\documentclass[conference,10pt,topmargin=.7in]{IEEEtran}
\usepackage[utf8]{inputenc}

\usepackage{hyperref}
\usepackage{verbatim} 

\usepackage{graphicx}
\usepackage{caption}
\usepackage{subcaption}
\usepackage{amssymb}
\usepackage{ textcomp } 
\usepackage{color}
\usepackage{enumerate}
\usepackage{mathrsfs}
\usepackage{multicol}
\usepackage{hyperref}
\usepackage{amsopn}

\usepackage{amsmath}
\usepackage{amsthm}
\usepackage{amssymb}

\usepackage{multicol}

\usepackage{algorithm}
\usepackage{algorithmic} 

\newcommand{\K}{\mathbb{K}}

\newcommand{\R}{\mathbb{R}}

\newcommand{\calA}{\mathcal{A}}

\newcommand{\calC}{\mathcal{C}}

\newcommand{\calN}{\mathcal{N}}

\newcommand{\calT}{\mathcal{T}}
\newcommand{\calU}{\mathcal{U}}
\newcommand{\calW}{\mathcal{W}}

\newcommand{\abs}[1]{\left\vert #1 \right\vert}
\newcommand{\norm}[1]{\Vert #1 \Vert}

\newcommand{\set}[1]{\left\lbrace #1\right\rbrace}

\newcommand{\erw}[1]{\mathbb{E}\left( #1 \right)}

\usepackage{dsfont}

\newcommand{\geqsim}{\gtrsim}

\DeclareMathOperator{\id}{id}

\newcommand{\argmin}{\mathop{\mathrm{argmin}}}

\newtheorem{lem}{Lemma}

\newtheorem{theo}[lem]{Theorem}

\theoremstyle{definition}

\numberwithin{lem}{section}

\usepackage{etoolbox}
\let\bbordermatrix\bordermatrix
\patchcmd{\bbordermatrix}{8.75}{4.75}{}{}
\patchcmd{\bbordermatrix}{\left(}{\left[}{}{}
\patchcmd{\bbordermatrix}{\right)}{\right]}{}{}

\title{Guaranteed blind deconvolution and demixing via hierarchically sparse reconstruction. }
\author{\IEEEauthorblockN{%
Axel Flinth\IEEEauthorrefmark{1}, %
Ingo Roth\IEEEauthorrefmark{2}\IEEEauthorrefmark{3}, %
Benedikt Gro\ss\IEEEauthorrefmark{4}, Jens Eisert\IEEEauthorrefmark{3}, Gerhard Wunder\IEEEauthorrefmark{4}} %
\IEEEauthorblockA{%
    \IEEEauthorrefmark{1}Institute for Electrical Engineering, Chalmers University of Technology, Gothenburg, Sweden \\ \IEEEauthorrefmark{2}Quantum Research Centre, Technology Innovation Institute, Abu Dhabi, UAE \\
    \IEEEauthorrefmark{3}Dahlem Center for Complex Quantum Systems, Freie Universit\"at Berlin, Berlin, Germany \\
    \IEEEauthorrefmark{4} Cybersecurity and AI group, Freie Universität Berlin, Berlin, Germany}
}

\date{\today}

\usepackage{algorithmic} 
\usepackage{algorithm} 
\usepackage{tikz}
\usepackage{xfrac} 

\usepackage{cite} 

\definecolor{ingo}{rgb}{1,.2,.4}

\definecolor{axel}{rgb}{0,.6,.8}

\definecolor{gerhard}{rgb}{1,.8,.1}

\usepackage{amsmath}

\begin{document}

\maketitle

\begin{abstract}
   The blind deconvolution problem amounts to reconstructing both a signal and a filter from the convolution of these two. It constitutes a prominent topic in mathematical and engineering literature. In this work, we analyze a sparse version of the problem: The filter $h\in \R^\mu$ is assumed to be $s$-sparse, and the signal $b \in \R^n$ is taken to be $\sigma$-sparse, both supports being unknown.  
   We observe a convolution between the filter and a linear transformation of the signal. 
   Motivated by practically important multi-user communication applications, we derive a recovery guarantee for the simultaneous demixing and deconvolution setting. 
   We achieve efficient recovery by relaxing the problem to a  hierarchical sparse recovery for which 
   we can build on a flexible framework. 
   At the same time, for this we pay the price of some sub-optimal guarantees compared to the number of free parameters of the problem.  
   The signal model we consider is sufficiently general to capture many applications in
   a number of engineering fields. 
   Despite their practical importance, 
   we provide first rigorous performance guarantees for efficient and simple algorithms for the bi-sparse and generalized demixing setting. 
   We complement our analytical results by presenting results of numerical simulations.
   We find evidence that the sub-optimal scaling $s^2\sigma \log(\mu)\log(n)$ of our derived sufficient condition is likely overly pessimistic and that the observed performance is better described by a scaling proportional to $ s\cdot \sigma$ up to log-factors. 
   
\end{abstract}

\section{Introduction}
The \emph{blind deconvolution problem} is the problem of recovering both a filter $h \in \R^\mu$ and a message $x \in \R^\mu$ from its convolution $y=h*x$. In the centre of this work are well-motivated 
circular convolutions of the form 
\begin{align*}
    [h*x]_i = \sum_{k \in[\mu]}h_k x_{i-k}\,,
\end{align*}
where $[\mu]$ is a shorthand for the set $\set{0, 1, \dots, \mu-1}$ of residual classes modulo $\mu$. This model naturally emerges within the context of wireless communications: When transmitting a message $x$ over a wireless channel, the signal are scattered on random features in the environment. This means that the signal arriving at the receiver is the superposition of damped and delayed copies of $x$ -- these effects are described by the convolution of $x$ with a channel filter $h$. 
Recovering the message $x$ at the receiver without knowledge of $h$ amounts to solving a blind deconvolution problem.

\begin{figure}[h]
    \centering
    \includegraphics[width=.43\textwidth]{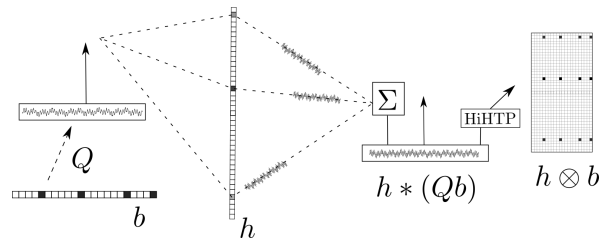}
    \caption{Our communication model. The transmitter translates a message $b$ to a sequence $Qb$ which then is sent over a channel. Due to delays and scattering, the receiver measures $h*(Qb)$. Thanks to the hierarchically sparse structure of $h\otimes b$, it can be recovered 
    using HiHTP.}
    \label{fig:blind_deconvolution}
\end{figure}

In this work, we consider
the $\emph{bi-sparse}$ version of the blind deconvolution problem. Concretely, we assume that the filter $h \in \R^\mu$ is $s$-sparse and that the message $x$ can be sparsely represented in a known dictionary $Q\in \R^{\mu,n}$, i.e., $x=Qb$ for some $s$-sparse vector $b \in \R^n$. The vector $b$ will be referred to as the \emph{signal}. 
This version of the problem is well-motivated in various communication scenarios: Scattering typically occurs only along a very small number of paths, such that $h$ is effectively sparse.
On a resource limited end device a typically sparse, compressible signal $b$ is encoded via an inexpensive linear transformation $Q$ and then transmitted.\smallskip

{\bf Hierarchical sparsity.} %
The bi-sparse deconvolution problem can be rewritten into a linear recovery problem of a \emph{hierarchically sparse vector} (as already observed in our earlier work
Ref.~\cite{wunder2018secure}). 
Since the blind convolution map $$C: \R^\mu \times \R^n \to \R^\mu, (h,b) \mapsto h*(Qb)$$ is bi-linear, there exists a unique linear map $\calC : \R^\mu \otimes \R^n \to \R^\mu$ with $$\calC(h \otimes b) = h*(Qb), \quad h \in \R^\mu, b \in \R^n,$$ here, $\otimes$ denoting the tensor product. 
Thus, in a `lifted version' the deconvolution problem becomes a linear inverse problem of a highly structured signal. 
In this work, we restrict our attention to the following structure.   

Now, if $h$ and $b$ are sparse, their tensor product $h \otimes b = \sum_{k \in [\mu]} h_k e_k \otimes b$, with the canonical basis $(e_k)_i = \delta_{k,i}$, can be interpreted as a vector that consists of   $\sigma$-sparse blocks $(h_k e_k \otimes b)_{k \in [\mu]}$ and only $s$ out of the $\mu$ blocks are  non-vanishing.  
Such a block-sparse vector with, in general distinct,  sparse blocks is called $(s,\sigma)$-(hierarchically)-sparse \cite{SprechmannEtAl:2010,FriedmanEtAl:2010, SprechmannEtAl:2011, SimonEtAl:2013}.

In Refs.~\cite{hiHTP, RothEtAl:2016:Proceedings}, it has been shown that such vectors can be efficiently reconstructed from linear measurements $\calA(h \otimes b)$ with the  HiHTP algorithm, the hierarchical hard-thresholding pursuit, Algorithm~\ref{alg:HiHTP}.  
The HiHTP is guaranteed to converge to the ground truth signal 
if the linear map $\calA$ exhibits the \emph{hierarchical restricted isometry property} (HiRIP). 
Concretely, we define the $(s,\sigma)$-HiRIP constant as 
\begin{align*}
    \delta_{(s,\sigma)}(A) = \sup_{u \, (s,\sigma)\text{-sparse}, \norm{u}=1} \abs{\norm{\calA(u)}^2  -\norm{u}^2}\, .
\end{align*}
If the $\delta_{(2s,3\sigma)}(\calA)$ is smaller than $\tfrac{1}{\sqrt{3}}$, the HiHTP succeeds at recovering any $(s,\sigma)$-sparse vector robustly against model-mismatch and stable against noise. 
The HiHTP algorithm is simple to implement and has a run time dominated by the costs of the matrix-vector multiplication, i.e., $\mathcal{O}(m\mu n)$ without further assumption. The projection step itself is efficient with time complexity $\mathcal{O}(\mu n)$. 
In addition, the `expensive' steps can be computed in parallel for each block. 
We refer to Ref.~\cite{eisert2021hierarchical} for a more complete introduction to hierarchical compressed sensing  and the line of work \cite{RothEtAl:2016:Proceedings, hiHTP, roth2018hierarchical,WunderEtAl:2018:Performance, Wunder2019_TWC, WunderEtAl:2018:Hierarchical, WunderEtAl:2017:HiHTP, wunder2018secure, RothEtAl:2020:Semidevicedependent, grosshierarchical,  flinth2021hierarchical}.

\renewcommand{\algorithmicrequire}{\textbf{Input:}}
\renewcommand{\algorithmicensure}{\textbf{Output:}}
\begin{algorithm}[b] 
\caption{HiHTP \label{alg:HiHTP}}
\begin{algorithmic}[1]
\REQUIRE{ vector $y\in\R^m$, measurement operator $\calA: \R^{\mu n} \to \R^m$, sparsity levels $s,\sigma$}
\STATE Init ${x}^{(0)} = 0$
\REPEAT
    \STATE $\bar{x}^{(t)} = x^{(t-1)} + \tau^{(t)} \calA^*\left(y-{\calA}{x}^{(t-1)}\right)$
    \STATE $I^{(t)} = \operatorname{support} 
    \argmin\limits_{x}
    \| \bar{x}^{(t)} - x\|$ s.t. $x$ $(s,\sigma)$-sparse
    \STATE ${x}^{(t)} = \argmin\limits_{{x}} \frac12 \|{y}-{\calA}{x}\|^2 \ $ s.t. $\  \operatorname{support}(x)\subseteq I^{(t)}$
\UNTIL{stopping criterion is met at $t = t^\ast$}
\ENSURE{$(s,\sigma)$-sparse vector $ x^{(t^\ast)}$}
\end{algorithmic}
\end{algorithm}

Thus, by relaxing the bi-sparse deconvolution problem to a hierarchically sparse recovery problem, we can use the HiHTP algorithm to solve it.  
In this work, we work out theoretical guarantees and identify parameter regimes in which we can ensure that this strategy succeeds. 
In particular, we establish the HiRIP for the blind convolution operator. 
Furthermore, we show that our hierarchical approach and  its guarantees can be straight-forwardly generalized to include the demixing of multiple observed convolutions. 
\smallskip

{\bf Previous work.} %
The blind deconvolution problem has a long history with an extensive body of literature. 
We, therefore, restrict our focus on works that specifically treat sparse versions of the problem.  Related treatises of non-sparse versions can be found, e.g., 
in Refs.~\cite{li2019rapid,ahmed2013blind,ling2017blind,jung2018blind}. 

A popular method for solving the bi-linear reconstruction method is via \emph{alternating minimization} \cite{NetrapalliAlternating,Bresler2015blind,lee2016blind}. Alternating minimization generally refers to alternately optimizing $\norm{C(h,b)-y}$ over $h$ and $b$ while leaving the respective other variable constant.
Since the convolution is linear in each argument, each subproblem is effectively a classical compressed sensing problem, and can be solved using a number of different techniques, e.g. iterative hard thresholding \cite{foucart2011hard} or CoSAMP \cite{needell2009cosamp}. 
    
For the alternating minimization approach, the authors of \cite{Bresler2015blind,lee2016blind} derive a recovery guarantee. 
This guarantee is however only applicable when $h$ and $x=Qb$ are \emph{spectrally flat} (somewhat more formally meaning that their Fourier transforms have entries of relatively equal magnitude). 
Crucially, this assumption is actively used in their algorithm: One step of their algorithm consists of projecting onto the set of spectrally flat signals, a step which is hard to perform exactly. The authors hence need to resort to heuristics for the projection. 
However, accepting this caveat, the authors prove convergence already when only observing $(s+\sigma)\log(\mu)^5$ of the entries in $h*x$, which is up to log-terms sample optimal  \cite{kech2017optimal,li2017identifiability}.

Lifted approaches only assuming sparsity of the message $b \in \R^n$ have been treated in Refs.~\cite{ling_2015,flinth2018sparse}. More specifically, $h$ is assumed to lie in an a priori known $s$-dimensional subspace of $\R^\mu$. Using the $\ell_1$- \cite{ling_2015} or the $\ell_{1,2}$-norm \cite{flinth2018sparse} as a regularizer, recovery can be guaranteed when $\mu \geqsim s\sigma \log(sn)\log(\mu)^2$. Since here only the $s\sigma$-sparse nature of the lifted vector $h \otimes b$ is used, this can also be viewed as a `pseudo-optimal' sampling complexity in this setting. 
The price for the relaxation to sparse signals is that the scaling $s+ \sigma$ is in principle not reachable. 
Compared to the setting considered in our work, the assumption that $h$ lies in a known subspace is a significant simplification.

The optimal scaling requires to also enforce a unit rank constraint in the lifted setting. 
One approach is to perform a gradient descent projected onto the set of (bi)-sparse \emph{and} low-rank matrices.
As is thoroughly discussed in Ref.~\cite{foucart2020jointly}, there is however no efficient algorithm to compute the projection onto the  set of sparse and low-rank matrices. A canonical way to circumvent this is to alternate between projections onto the two sets. This approach is for instance investigated in Ref.~\cite{eisenmann2021riemannian}. 
There, a local convergence guarantee is presented under optimal sample complexity,  however only under a fully Gaussian measurement model neglecting the structure of the blind deconvolution problem. 
Similar results are given in Ref.~\cite{lee2017near} -- their guarantee is however only sample optimal under an additional assumption on the signal. 
In this context, Ref.~\cite{bahmani2016near} should also be mentioned -- in there, a global convergence in just two alternations steps is shown. This work however assumes a nested measurement structure tailor-made for a jointly low-rank and sparse setting, which is not applicable in our setting. 
In the light of these approaches, our relaxation to hierarchically sparse signals can been seen as the closest structure for which the projection is efficient. 
\smallskip 
 
 {\bf Outline.} %
 In Sec.~\ref{sec:main_result}, we present and discuss our theoretical guarantee for recovering $h \otimes b$ from the blind convolution measurement $\calC(h \otimes b)$. In Sec.~\ref{sec:demixing}, we describe how the hierarchical framework can be used to easily translate our results to a multi-user setting, where a blind deconvolution and demixing-problem arises. In Sec.~\ref{sec:numerics}, we study the scaling behaviour in  numerical simulations.
 
\section{Main Result} \label{sec:main_result}
Let us begin by presenting our measurement model more thoroughly. 
As outlined in the introduction, our aim is to recover the lifted filter-message tensor $h \otimes b \in \R^{\mu} \otimes \R^n$ from the measurement $$y= \calC(h \otimes b) = h*(Qb),$$ involving a circular convolution. 
Our recovery guarantee relies on a particular model for the matrix $Q$.
\smallskip

{\bf Random model for $Q$.} We assume that $Q$ can be decomposed as $Q=UA$. 
Here with suitable $m$, the operator $A \in \R^{m,n}$ is a matrix with small standard RIP constant  $\delta_{\sigma}$, and $U \in \R^{\mu,m}$ is an isotropically normalized Gaussian matrix, i.e., the entries of $U$ are independent and $\calN(0,\mu^{-1})$ distributed.

Note that $A=\id$ is a very viable choice, but the ability to choose it as a standard `compressed sensing matrix' is interesting both theoretically, and from a practitioners standpoint, e.g. in the communication setup.  
We can interpret $A$ as a codebook, which maps the message $b$ to a `codeword' $Ab \in \R^m$. 
This codeword vector is subsequently converted into a sequence $UAb \in \R^\mu$ and sent over the channel to the receiver.
We can now state our main result.
\begin{theo} \label{th:main_result}
Fix $\delta_0 \in (0,1)$ and let $\epsilon>0$. Further assume that 
\begin{align}
    \mu \geqsim (s^2\log(\mu) &+ s^2\sigma\log(n))  \cdot \delta_0^{-2} \cdot  \max(1, \log(\epsilon^{-1})), \label{eq:samplecomp}
\end{align}
where $\geqsim$ means that the inequality needs to hold up to a universal constant. We then have
\begin{align*}
    \delta_{(s,\sigma)}(\calC) \leq (1+\delta_\sigma(A))^2\delta_0  + \delta_{\sigma}(A)
\end{align*}
with a probability at least $1-\epsilon$. 
\end{theo}

As a direct consequence, we establish that the condition \eqref{eq:samplecomp} is sufficient to ensure that, with probability of at least $1-\epsilon$, the HiHTP algorithm~\ref{alg:HiHTP} succesfully recovers each $(s,\sigma)$-sparse ground truth $h \otimes b$ by
Ref.~\cite[Thm.~1]{hiHTP}.
\smallskip

{\bf Discussion.} Disregarding logarithmic terms, and terms related to adjustable threshold, our complexity bound scales as $s^2\sigma$. This is obviously not even close to the sample optimal number of measurements  $s+\sigma$. 
Most of this loss in optimality is expected in any approach that only considers the hierarchically sparse structure of the problem. 
In particular, as discussed in the literature review, an optimal scaling can only be achieved when the bisparse and low-rank structure is explicitly taken into account. In contrast, the HiHTP-algorithm would work just as good when fed with a signal $(w_1, \dots, w_n)$ where the blocks are neither equal nor share a common support. 
However, the existing approaches in the literature require additional assumptions to establish recovery guarantees and rely on heuristics for hard projection steps. 
The hierarchically sparse structure, that allows for an efficient projection step, would instead suggest a number of $s\sigma$ measurement from looking at the number of free parameters. 
Compared to this scaling, the derived sampling complexity still has an additional factor of $s$. 
This might be an artefact of the proof techniques employed. 
Available guarantees in the literature meeting this sample complexity so far relied on a priori knowledge of the support of either $h$ or $b$. 
Thus, although the sample complexity of our result is neither optimal nor `pseudo-optimal', it still significantly complements the state-of-the-art.

We suspect that the HiHTP algorithm in fact reaches the `pseudo-optimal' scaling of $s\sigma$ in the blind-deconvolution problem. 
A small numerical study supporting this claim is given in Sec.~\ref{sec:numerics}.
\smallskip

{\bf Proof sketch.} We provide a sketch of the proof of Thm.~\ref{th:main_result}, concentrating on the case $A=\id$. The entity we need to bound is
\begin{align*}
    \sup_{\substack{w  \, (s,\sigma)\text{-sparse},  \\ \norm{w}_2=1}} \abs{\norm{\calC(w)}^2 - \norm{w}^2}.
\end{align*}
Through direct calculation, one can relate 
\begin{align*}
        \norm{\calC(w)}^2 =  \sum_{i,j\in [\mu]} \sum_{r,s \in [n]} \gamma_{r,i}\gamma_{s,j}  \calW_{(i,r),(s,j)} := T_\calW,
    \end{align*}
    where $\gamma_{r,i}$ are independent, centred Gaussians with variance $1$, and 
$\calW \in (\R^{n,n})^{\mu,\mu}$ is the Block Toeplitz matrix
    \begin{align}
        \calW = (W^{i-j})_{i,j\in [\mu]}, \quad W^\ell = \tfrac{1}{\mu} \sum_{k\in [\mu]}w_k w_{k+\ell}^*\,. \label{eq:block_toeplitz}
    \end{align}
We denote by $\calT_{s,\sigma}$ the set of such Block Toeplitz matrices which are generated through \eqref{eq:block_toeplitz} as $w$ traverses the set of $(s,\sigma)$-sparse, normalized vectors. 
Since $\erw{\norm{\calC(w)}^2} =  \norm{w}^2$, the entity we need to bound is hence equal to 
\begin{align*}
    \sup_{\calW \in \calT_{s,\sigma}} \abs{T_\calW - \erw{T_\calW}}
\end{align*}
This is a supremum of a random centered process. The technique of identifying RIP constants as such suprema is well-established in the compressed sensing literature, see e.g.,~Refs.~\cite{FouRau2013,KrahmerSuprema}. 

In order to bound the supremum with high probability, we utilize the ideas of \emph{generic chaining} \cite{genericChaining}. In essence, this framework tells us that if we can bound all increments $(\calT_\calU - \calT_\calW)$, for $\calU, \calW \in \calT_{s,\sigma}$, a bound for the supremum follows. 
We can achieve such increment bounds by subdividing $\calT_\calW$ into two processes, whose
 increments form second-order Gaussian chaos \cite{genericChaining}, and sums of independent subexponential variables, respectively. Both these types of random processes enjoy well-established concentration inequalities -- for the latter, we in particular apply the Bernstein inequality \cite{vershynin2010introduction}.
 
We then translate the concentration results for the increments into a bound on the suprema per se by estimating the `size' of $\calT_{s,\sigma}$ -- formally, we estimate the set's so-called $\gamma$-functionals of Talagrand. 
We bound these by invoking theory on Block Toeplitz matrices \cite{Bottcher1999} that allows us to relate them to the $\gamma$-functionals of the set of $(s,\sigma)$-sparse vectors. 
The latter can be estimated with well-known techniques. Putting everything together, one deduces that \eqref{eq:samplecomp} is sufficient to guarantee the HiRIP with high probability.

The generalization to $A \neq \id$ follows by first relating the  standard restricted isometry constant $\delta_\sigma(\calC)$ with the one of $A$, and the  RIP-constant  of $\widehat{\calC}:\R^\mu \otimes \R^m \to \R^\mu$, defined through $\widehat{\calC}(h,v) = h*(Uv)$, however, 
restricted block vectors of the form $(Av_1, \dots Av_\mu)$ with $v\in T_{s,\sigma}$ instead of $T_{s,\sigma}$ directly. Since $A$ acts almost isometrically on the latter 
set, the rest of the proof proceeds as 
above, with minor modifications.

Needless to say, each of the steps outlined above is rather technical, and a full proof can not be presented here in detail. 
We postpone the detailed proof to an upcoming journal version of this work.
\smallskip

\section{Multiuser case} \label{sec:demixing}
Above, we utilized that $h \otimes b$ is $(s,\sigma)$-sparse, i.e., hierarchical sparsity in two levels in order to solve a bisparse blind deconvolution problem. 
One of the major strength of the hierarchical strategy is that it can very flexibly incorporate more complicated settings where, e.g. multiple convoluted signals are linearly superimposed. 
Hierarchical compressed sensing naturally extends to deeper hierarchies of sparsity levels \cite{eisert2021hierarchical,hiHTP}. 
For instance, we say that a block vector $(X_1, \dots, X_N)$ consisting of $S$ non-vanishing blocks that itself are $(s,\sigma)$-sparse is $(S,s,\sigma)$. 
The HiHTP algorithm can still be employed for the recovery of such signals when using the equally efficient projection onto  $(S,s,\sigma)$-sparse vectors in line $4$. 
Recovery can again be guaranteed with the help of a hierarchically restricted isometry property \cite{hiHTP}.

Three-level hierarchically sparse vectors, naturally arise in a \emph{sparse blind demixing and blind deconvolution} problem. 
This problem consists in recovering a set of filter-signal pairs $(h_i,b_i)$ from observations of $M$ mixtures of their convolutions, i.e.,
\begin{align}
    y_j = \sum_{i \in [N]} d_{j,i} h_i * Q_ib_i,\quad j \in [M] \label{eq:mixing}
    \,.
\end{align}
Note that we do not necessarily assume that the matrices $Q_i$, $i \in [N]$ are equal. If we assume that only $S$ of the message-filter vectors $h_i \otimes b_i$ are nonzero, and that each of these are $(s,\sigma)$-sparse as before, the collection of filter-message vectors $\sum_{i \in [N]} e_i \otimes h_i \otimes b_i$ is $(S,s,\sigma)$-sparse.

\begin{figure}
    \centering
    \includegraphics[width=.38\textwidth]{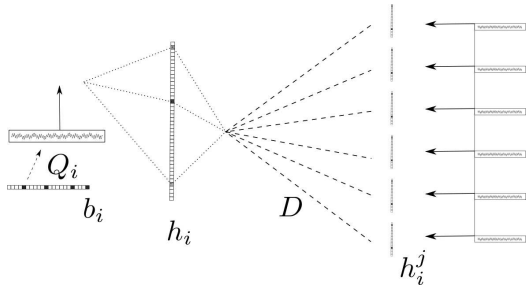}
    \caption{The multiantenna model. The filters of one users for different antenna are correlated. The correlation is described by the matrix $D$.}
    \label{fig:demixing}
\end{figure}

\begin{figure}
    \centering
    \includegraphics[width=.38\textwidth]{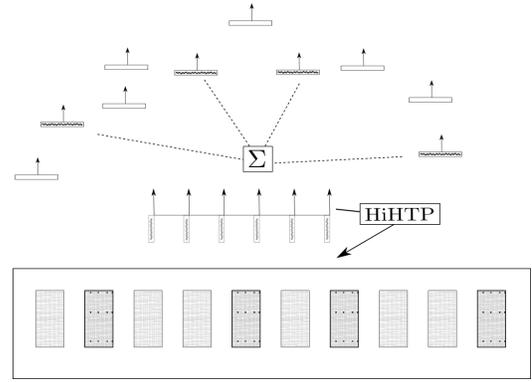}
    \caption{ The sparse multi-user model. The multi-antenna reciever can de-entangle the individual $h_i \otimes b_i$ from the collective measurements due to the three-level hierarchical sparsity.}
    \label{fig:demixing}
\end{figure}
\smallskip

{\bf A multi-user model.} Equation \eqref{eq:mixing} can be used to model a multi-user, multi-antenna communication scenario. Imagine $N$ users simultaneously transmitting signals, as in the previous sections, to a receiver with $M$ antennas. 
For each pair of a user and an antenna at the receiver, there will be a unique filter $h^j_i \in \K^\mu$, $j \in [M], i \in [N]$. For each user, we assume that the filters are linearly correlated in the following sense: 
For each user $i$, there exists a `basic' filter $h_i \in \R^\mu$ so that the other filters of the user are given through
$    h^{j}_i = d_{j,i} h_i$ 
for some scalars $d_{j,i}$.

This assumption can be motivated as follows: Let us imagine a wavefront that at time $t$ results in a response $v(t) \in \R$ in one of the antennas. Due to relative path differences to the other antennas, it will result in a collective response $d(\theta)v(t) \in \R^M$ in all antennas for some function $d: \Omega \to \R^M$, where $\Omega$ is a set of angles. Consequently, if  the scattered transmitted signals of one user arrives with delays $k$ from directions $\theta_k$, the response of the  antenna $q$ at time $t$ will be 
 \begin{align*}
     y_j(\ell)  = \sum_{k \in [\mu]} h_i(k) Q_ib_i(\ell-k) d_j(\theta_k).
 \end{align*}
 If now for each user, \emph{the wavefronts are arriving from the same angle $\theta^i$}, this reduces to 
 \begin{align*}
      y_j(\ell) &=  \sum_{k \in [\mu]} h_i(k) Q_ib_i(\ell-k) d_j(\theta^i) \\
      &= [(d_j(\theta^i)h_i)*(Q_i b_i)](\ell)\,. 
 \end{align*}
We, thus, find exactly the setting above with $d_{j,i}=d_j(\theta^i)$.

Now, adhering to our previous assumptions in the single user communication setting, each tensor $h_i\otimes b_i$ is $(s,\sigma)$-sparse. 
In a multi-user setting with a \emph{sporadic user activity}, as e.g. motivated by the internet of things,  in addition,  at each instance in time, only $S$ of the $N$ users are transmitting. %
Hence, we find that $\sum_{i \in [N]} e_i \otimes h_i \otimes b_i$ is $(S,s,\sigma)$-sparse. 
\smallskip

{\bf Theoretical guarantee.} The measurement model  \eqref{eq:mixing} defines a so called \emph{hierarchical measurement operator}  \cite{grosshierarchical,flinth2021hierarchical}. 
To be concrete, if we define $D=[d_0, \dots d_{N-1}] = (d_{j,i})_{j \in [M], i \in [N]}$, the collective measurement $y \in \R^M \otimes \R^\mu$ of all antennas is given by
\begin{align*}
    y = \sum_{i \in [N]}d_i \otimes (h_i * Q_ib_i)\in \R^{M}\otimes \R^\mu
\end{align*} 
 For such a hierarchical measurement operator,  Ref.~\cite[Theorem 2.1]{flinth2021hierarchical} states that if both the `mixing matrix' $D$  has the $S$-sparse RIP and each blind convolution operator $\calC_i: \R^\mu \otimes \R^n \to \R^\mu$ has the $(s,\sigma)$-sparse HiRIP, the entire operator \eqref{eq:mixing} has the $(S,s,\sigma)$-sparse HiRIP. 
 We get following corollary. 
\begin{theo}
 Assume that the  $D= (d_{i,j})_{i \in [M], j \in [N]}$ has an $S$-sparse RIP constant $\delta_S(D)<1$. Further assume that each  each blind convolution operator
 \begin{align*}
     \calC_i(h \otimes b) = h_i \otimes Q_i b
 \end{align*}
 obeys $\delta_{(S,s,\sigma)}(\calC_i)< \delta$. Then, the $(S,s,\sigma)$-HiRIP constant of the measurement \eqref{eq:mixing} is dominated by $\delta_S(D) + \delta+  \delta_S(D)  \cdot \delta$.
\end{theo}
The above proposition in combination with Theorem \ref{th:main_result} in particular proves that we can simultaneously recover $S$ active $\sigma$-sparse filters $h_i\in \R^{\mu}$ and $s$-sparse messages $b_i\in \R^n$ from\linebreak $M \sim S\log(N)$ mixtures of the form \eqref{eq:mixing}, provided $\mu~\geqsim~s^2(\log(\mu)+\sigma \log(n))\log(\mu)$.
\smallskip

\section{Numerical experiments} \label{sec:numerics}
We complement our analytical guarantees with a brief numerical simulations. 
In particular, we want to investigate whether the quadratic scaling in $s$ of our main results is an  artifact of the proof technique. 
\smallskip

{\bf Details of implementation.} We have implemented the HiHTP algorithm using the python package \texttt{PyTorch} --  facilitating parallel computations on the GPU. 
Our implementation assumes that the application of the linear operator $Q: \R^n \to \R^\mu$ and its dual $Q^*:\R^\mu \to \R^n$ are capsuled.
In particular, we do not assume that their matrix representations are available,
reducing memory requirements and making it possible to utilize fast matrix-vector multiplications if available.
\smallskip

{\bf Experimental setup.} In all of the experiments, we set $A=\id$ and in particular $m=n$. We choose $Q=U$ as  a properly renormalized standard Gaussian matrix. 
We try to solve instances of the blind deconvolution problem for different values of $s, \sigma $ and $\mu$. 

\begin{table}[b] 
    \centering
    \begin{tabular}{c|| c | c | c | }
        $n \backslash \sigma$ & $5$ & $10$ & $15$  \\
        \hline 50 & $10,20, \dots , 120$ & $10,20, \dots , 120$ & $10,20, \dots , 120$ \\
         350 & $10,20, \dots,  120$ &  $ 20,50, \dots, 350$ & $20,50, \dots, 350$
    \end{tabular}
    \caption{The ranges for $\mu$ for the different experiments.}
    \label{tab:mu_values}
\end{table}

The ranges of values of the sparsity parameters 
are given by $\sigma =5,10,15$ and $s = 1, \dots, 7$. We test two values for $n$, $n=50$ and $n=350$. The number of measurements tested is not the same for different values of $\sigma$ and $n$ and given in Table~\ref{tab:mu_values}. The ranges have manually been chosen to capture the phase transition for each setting. 
For each quadruple $(n,\mu,s,\sigma)$, we draw sparse vectors $b$ and $h$ at random. 
The vector $b$ is constructed by choosing a $\sigma$-sparse support uniformly at random, and fill the non-zero positions with Rademacher variables ($\pm 1$,  with equal probability). 
The filter $h$ is constructed with uniformly at random $s$-sparse support and non-vanishing entries independently drawn from $\calN(0,1)$. For each data point, we perform $100$ experiments. In the cases when $\mu<s\cdot \sigma$, we declare a failure preemptively, since we will not be able to recover the signal even if we pinpoint the correct support anyway. The HiHTP-algorithm is halted after at most $10$ iterations, and a success is declared when the final relative error is smaller than $10^{-4}$.

\begin{figure}
    \includegraphics[width=.25\textwidth]{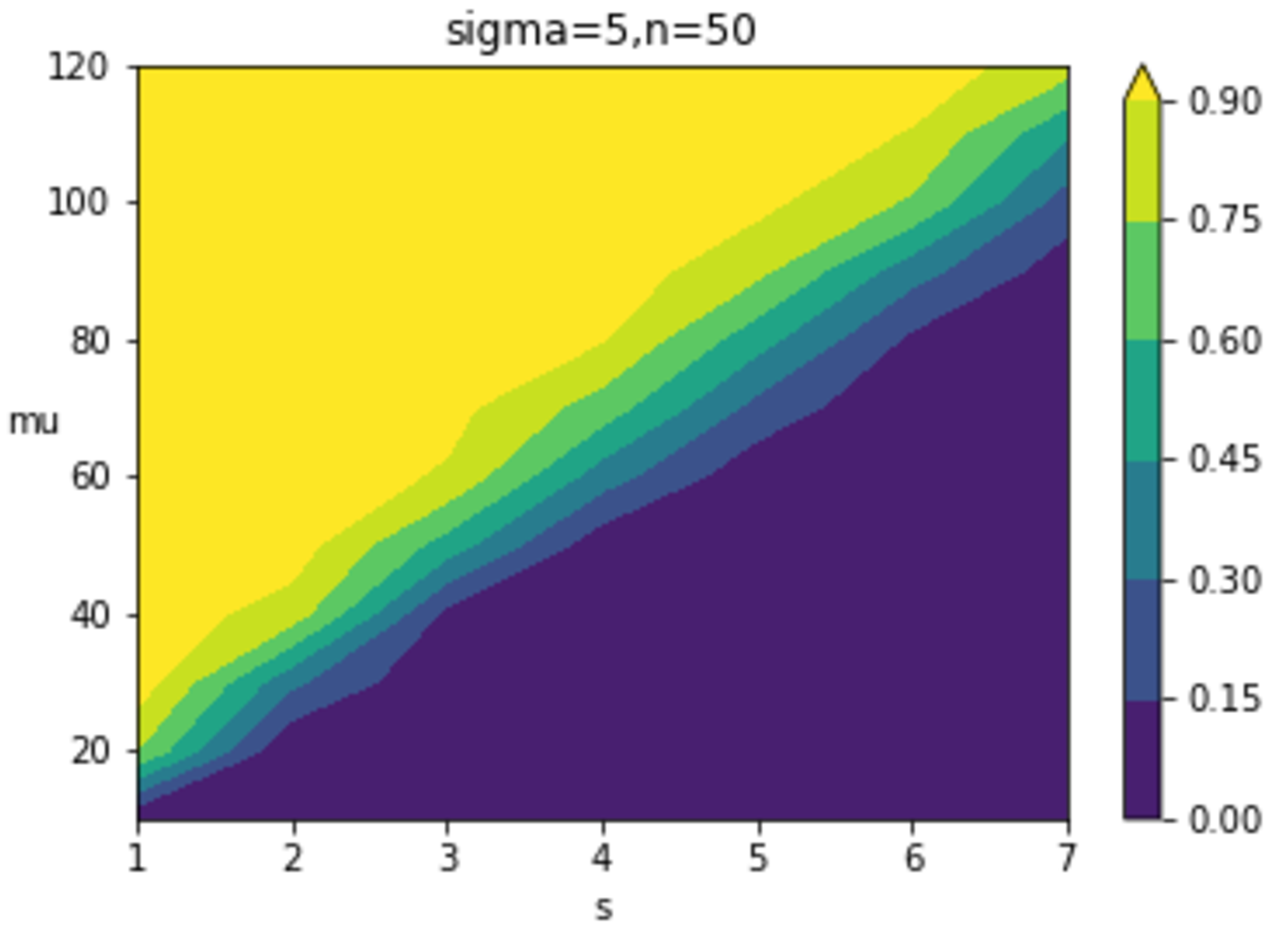}\includegraphics[width=.25\textwidth]{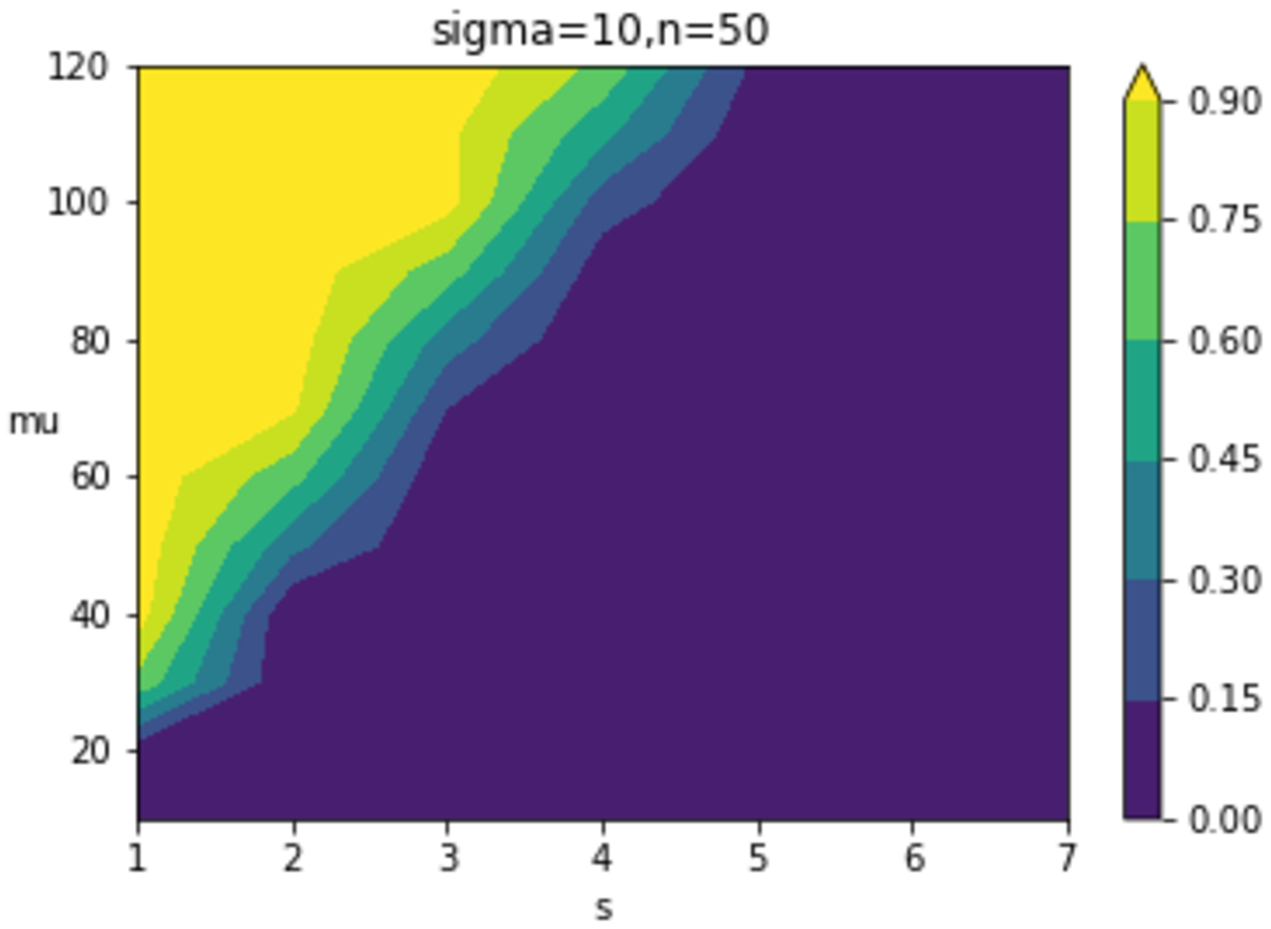}
    
    \includegraphics[width=.25\textwidth]{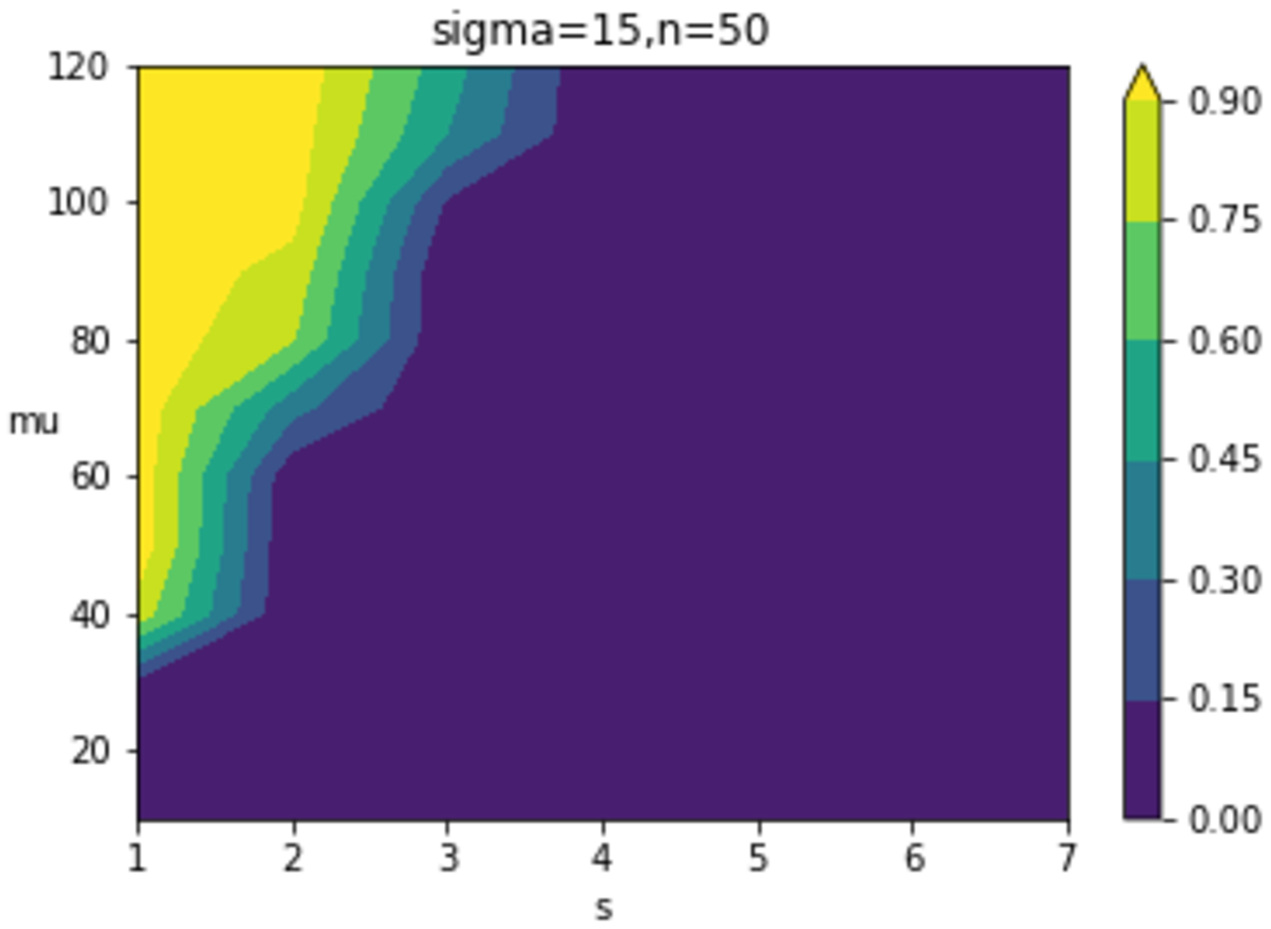}
    \caption{Results of the experiments for $n=50$. Shown is the fraction of successful experiments.}
    \label{fig:n=50}
\end{figure}

\begin{figure}
    \includegraphics[width=.25\textwidth]{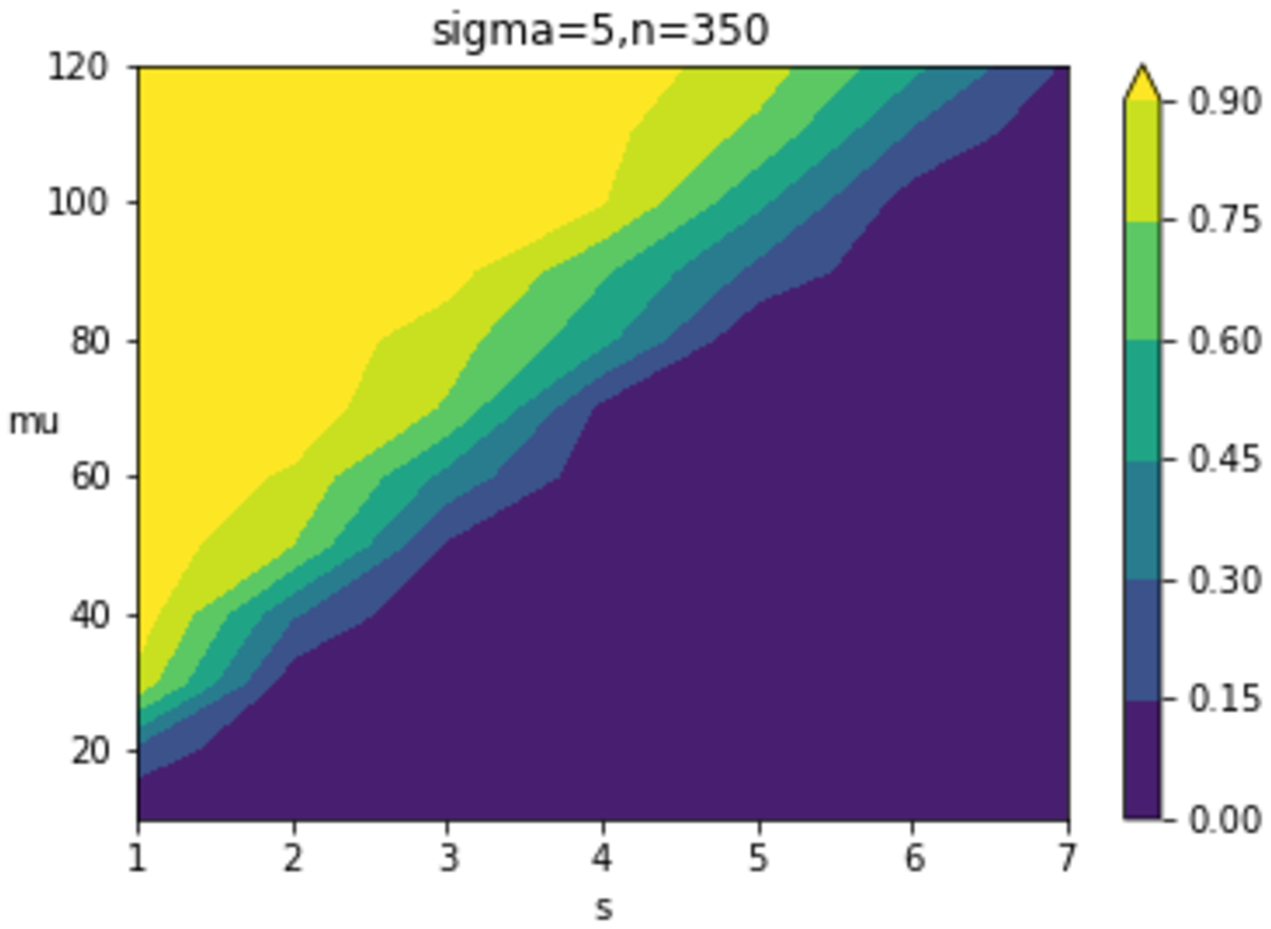}\includegraphics[width=.25\textwidth]{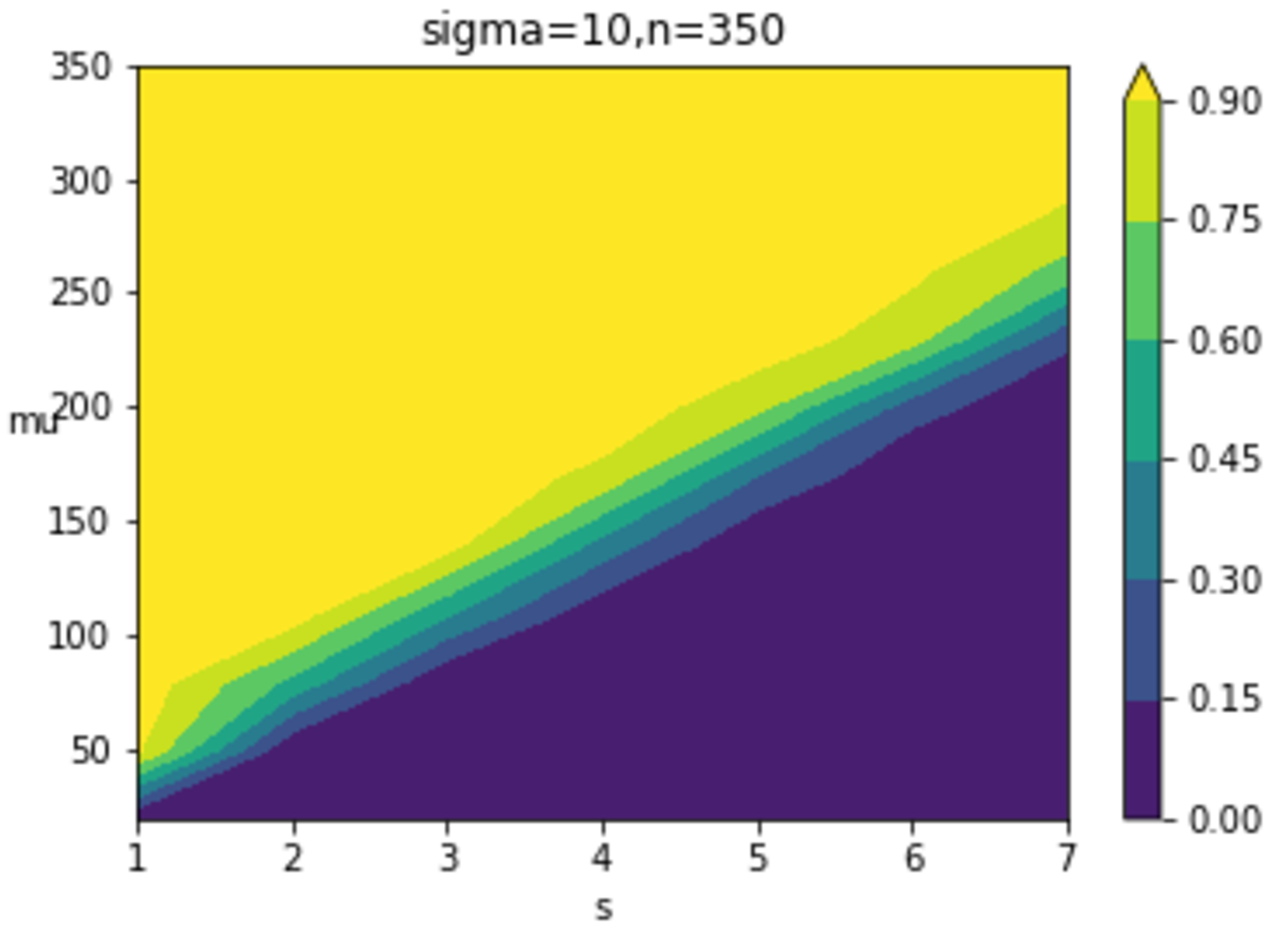}
    
    \includegraphics[width=.25\textwidth]{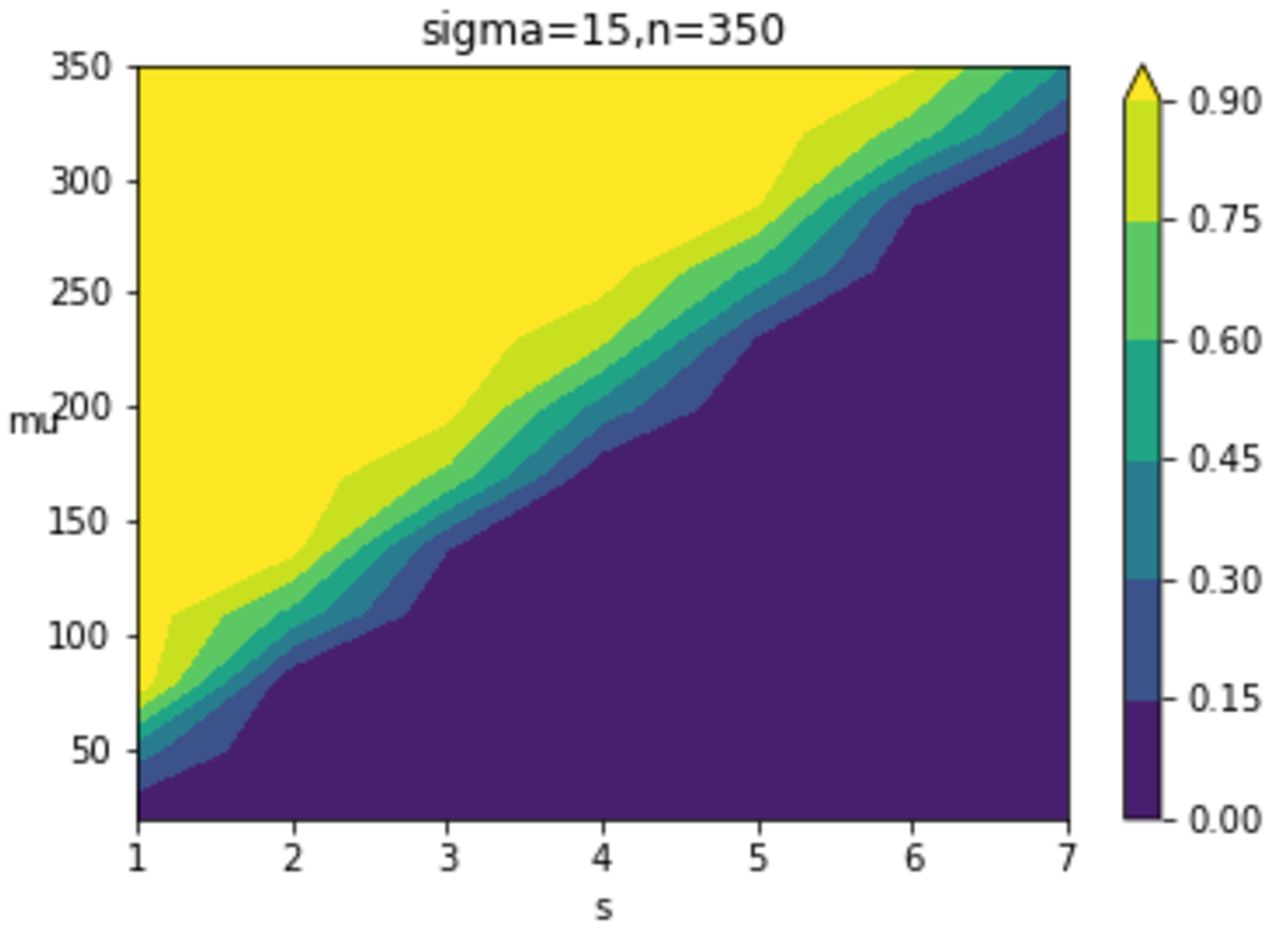}
    \caption{Results of the experiments for $n=350$. Shown is the fraction of successful experiments for the depicted parameter regime. Note the different $\mu$-range are different for the bigger $\sigma$-values.}
    \label{fig:n=350}
\end{figure}
\smallskip

{\bf Results.}  The results are depicted in Figures \ref{fig:n=50} and \ref{fig:n=350}. The figures clearly suggest, for all values of $\sigma$ and $n$, that the actual required sampling complexity in $s$ is linear, and not quadratic, in contrast to our main result. 
We furthermore see that the dependence on $n$ is relatively mild. 
The quadratic scaling seemingly only being a proof artifact increases the practical potential of our approach. However, it also means that our main result can probably be improved. 
We leave it to future work to close the gap between theory and practice.

\section*{Conclusion}
We have investigated the hierarchical compressed sensing as a means for solving the sparse blind deconvolution problem and generalizations thereof. 
This relaxation can be regarded as the closest structure to the original problem for which an efficient projection exists. 
We have derived a theoretical recovery guarantee for efficient, simple hard-thresholding algorithm, both for the blind deconvolution and also for the blind deconvolution and demixing problem, based on generalized notions of the restricted isometry property. In contrast, to existing result we do not rely on additional assumptions or heuristic approximations of the projection step. 
The flexibility and numerical efficiency of hierarchical thresholding together with the rigorous guarantees makes it a valuable candidate for many applications with stringent resource requirements. 
We suspect that the derived sample complexity is still not optimal for the hierarchical approach and established numerical evidence of an improved performance in practice.

    \bibliographystyle{plain}

\bibliography{block_model}

\begin{thebibliography}{10}

\bibitem{ahmed2013blind}
A.~Ahmed, B.~Recht, and J.~Romberg.
\newblock Blind deconvolution using convex programming.
\newblock {\em IEEE Trans. Inf. Th.}, 60:1711--1732, 2013.

\bibitem{bahmani2016near}
S.~Bahmani and J.~Romberg.
\newblock Near-optimal estimation of simultaneously sparse and low-rank
  matrices from nested linear measurements.
\newblock {\em Information and Inference: A Journal of the IMA}, 5:331--351,
  2016.

\bibitem{Bottcher1999}
A.~B{\"o}ttcher and B.~Silbermann.
\newblock {\em Introduction to Large Truncated Toeplitz Matrices}.
\newblock Springer New York, New York, NY, 1999.

\bibitem{eisenmann2021riemannian}
H.~Eisenmann, F.~Krahmer, M.~Pfeffer, and A.~Uschmajew.
\newblock Riemannian thresholding methods for row-sparse and low-rank matrix
  recovery.
\newblock {\em arXiv preprint arXiv:2103.02356}, 2021.

\bibitem{eisert2021hierarchical}
J.~Eisert, A.~Flinth, B.~Gro{\ss}, I.~Roth, and G.~Wunder.
\newblock Hierarchical compressed sensing.
\newblock {\em arXiv preprint arXiv:2104.02721}, 2021.

\bibitem{flinth2018sparse}
A.~Flinth.
\newblock Sparse blind deconvolution and demixing through
  $\ell_{1,2}$-minimization.
\newblock {\em Advances in Computational Mathematics}, 44, 2018.

\bibitem{flinth2021hierarchical}
A.~Flinth, B.~Gro{\ss}, I.~Roth, J.~Eisert, and G.~Wunder.
\newblock Hierarchical isometry properties of hierarchical measurements.
\newblock {\em arXiv preprint. arXiv:2005.10379}, 2021.

\bibitem{foucart2011hard}
S.~Foucart.
\newblock Hard thresholding pursuit: an algorithm for compressive sensing.
\newblock {\em SIAM Journal on Numerical Analysis}, 49:2543--2563, 2011.

\bibitem{foucart2020jointly}
S.~Foucart, R.~Gribonval, L.~Jacques, and H.~Rauhut.
\newblock Jointly low-rank and bisparse recovery: Questions and partial
  answers.
\newblock {\em Analysis and Applications}, 18:25--48, 2020.

\bibitem{FouRau2013}
S.~Foucart and H.~Rauhut.
\newblock {\em A Mathematical Introduction to Compressive Sensing}.
\newblock Birkhäuser, 2013.

\bibitem{FriedmanEtAl:2010}
J.~Friedman, T.~Hastie, and R.~Tibshirani.
\newblock A note on the group lasso and a sparse group lasso.
\newblock {\em Preprint}, 2010.
\newblock arXiv: 1001.0736.

\bibitem{grosshierarchical}
B.~Gro{\ss}, A.~Flinth, I.~Roth, J.~Eisert, and G.~Wunder.
\newblock Hierarchical sparse recovery from hierarchically structured
  measurements with application to massive random access.
\newblock In {\em 2021 IEEE Statistical Signal Processing Workshop (SSP)},
  pages 531--535, 2021.

\bibitem{jung2018blind}
P.~Jung, F.~Krahmer, and D.~St\"{o}ger.
\newblock Blind demixing and deconvolution at near-optimal rate.
\newblock {\em IEEE Trans. Inf. Th.}, 64:704--727, 2018.

\bibitem{kech2017optimal}
M.~Kech and F.~Krahmer.
\newblock Optimal injectivity conditions for bilinear inverse problems with
  applications to identifiability of deconvolution problems.
\newblock {\em SIAM Journal on Applied Algebra and Geometry}, 1:20--37, 2017.

\bibitem{KrahmerSuprema}
F.~Krahmer, S.~Mendelson, and H.~Rauhut.
\newblock Suprema of chaos processes and the restricted isometry property.
\newblock {\em Communications on Pure and Applied Mathematics}, 67:1877--1904,
  2014.

\bibitem{Bresler2015blind}
K.~Lee, Y.~Li, M.~Junge, and Y.~Bresler.
\newblock Stability in blind deconvolution of sparse signals and reconstruction
  by alternating minimization.
\newblock In {\em International Conference on Sampling Theory and Applications
  (SampTA)}, pages 158--162, 2015.

\bibitem{lee2016blind}
K.~Lee, Y.~Li, M.~Junge, and Y.~Bresler.
\newblock Blind recovery of sparse signals from subsampled convolution.
\newblock {\em IEEE Trans. Inf. Th.}, 63:802--821, 2016.

\bibitem{lee2017near}
K.~Lee, Y.~Wu, and Y.~Bresler.
\newblock Near-optimal compressed sensing of a class of sparse low-rank
  matrices via sparse power factorization.
\newblock {\em IEEE Trans. Inf. Th.}, 64:1666--1698, 2017.

\bibitem{li2019rapid}
X.~Li, S.~Ling, T.~Strohmer, and K.~Wei.
\newblock Rapid, robust, and reliable blind deconvolution via nonconvex
  optimization.
\newblock {\em Applied and computational harmonic analysis}, 47:893--934, 2019.

\bibitem{li2017identifiability}
Y.~Li, K.~Lee, and Y.~Bresler.
\newblock Identifiability and stability in blind deconvolution under minimal
  assumptions.
\newblock {\em IEEE Trans. Inf. Th.}, 63:4619--4633, 2017.

\bibitem{ling_2015}
S.~Ling and T.~Strohmer.
\newblock Self-calibration and biconvex compressive sensing.
\newblock {\em Inverse Problems}, 31:115002, sep 2015.

\bibitem{ling2017blind}
S.~Ling and T.~Strohmer.
\newblock Blind deconvolution meets blind demixing: Algorithms and performance
  bounds.
\newblock {\em IEEE Trans. Inf. Th.}, 63:4497--4520, 2017.

\bibitem{needell2009cosamp}
D.~Needell and J.~A. Tropp.
\newblock Cosamp: Iterative signal recovery from incomplete and inaccurate
  samples.
\newblock {\em Applied and computational harmonic analysis}, 26:301--321, 2009.

\bibitem{NetrapalliAlternating}
P.~Netrapalli, P.~Jain, and S.~Sanghavi.
\newblock Phase retrieval using alternating minimization.
\newblock {\em IEEE Trans. Sign. Proc.}, 63:4814--4826, 2015.

\bibitem{roth2018hierarchical}
I.~Roth, A.~Flinth, R.~Kueng, J.~Eisert, and G.~Wunder.
\newblock Hierarchical restricted isometry property for {K}ronecker product
  measurements.
\newblock In {\em 2018 56th Annual Allerton Conference on Communication,
  Control, and Computing (Allerton)}, pages 632--638. IEEE, 2018.

\bibitem{hiHTP}
I.~{Roth}, M.~{Kliesch}, A.~{Flinth}, G.~{Wunder}, and J.~{Eisert}.
\newblock Reliable recovery of hierarchically sparse signals for gaussian and
  kronecker product measurements.
\newblock {\em IEEE Transactions on Signal Processing}, 68:4002--4016, 2020.

\bibitem{RothEtAl:2016:Proceedings}
I.~{Roth}, M.~{Kliesch}, G.~{Wunder}, and J.~{Eisert}.
\newblock Reliable recovery of hierarchically sparse signals.
\newblock In {\em Proceedings of the third "International Traveling Workshop on
  Interactions between Sparse models and Technology" (iTWIST'16)}, pages
  36--38, 2016.

\bibitem{RothEtAl:2020:Semidevicedependent}
I.~Roth, J.~Wilkens, D.~Hangleiter, and J.~Eisert.
\newblock Semi-device-dependent blind quantum tomography.
\newblock {\em Preprint}, 2020.
\newblock arXiv:2006.03069.

\bibitem{SimonEtAl:2013}
N.~Simon, J.~Friedman, T.~Hastie, and R.~Tibshirani.
\newblock A {sparse}-{group} {Lasso}.
\newblock {\em J. Comp. Graph. Stat.}, 22:231--245, 2013.

\bibitem{SprechmannEtAl:2010}
P.~Sprechmann, I.~Ramirez, G.~Sapiro, and Y.~Eldar.
\newblock Collaborative hierarchical sparse modeling.
\newblock In {\em 2010 44th {Annual} {Conference} on {Information} {Sciences}
  and {Systems} ({CISS})}, pages 1--6, 2010.

\bibitem{SprechmannEtAl:2011}
P.~Sprechmann, I.~Ramirez, G.~Sapiro, and Y.~C. Eldar.
\newblock C-{HiLasso}: {A} {collaborative} {hierarchical} {sparse} {modeling}
  {framework}.
\newblock {\em IEEE Trans. Sig. Proc.}, 59:4183--4198, 2011.

\bibitem{genericChaining}
M.~Talagrand.
\newblock {\em The generic chaining. Upper and Lower Bounds of Stochastic
  Processes.}
\newblock Springer, 2005.

\bibitem{vershynin2010introduction}
R.~Vershynin.
\newblock Introduction to the non-asymptotic analysis of random matrices.
\newblock {\em arXiv preprint arXiv:1011.3027}, 2010.

\bibitem{WunderEtAl:2018:Hierarchical}
G.~Wunder, I.~Roth, M.~Barzegar, A.~Flinth, S.~Haghighatshoar, G.~Caire, and
  G.~Kutyniok.
\newblock Hierarchical sparse channel estimation for massive mimo.
\newblock In {\em WSA 2018; 22nd International ITG Workshop on Smart Antennas},
  pages 1--8. VDE, 2018.

\bibitem{WunderEtAl:2017:HiHTP}
G.~{Wunder}, I.~{Roth}, R.~{Fritschek}, and J.~{Eisert}.
\newblock Hihtp: A custom-tailored hierarchical sparse detector for massive
  mtc.
\newblock In {\em 2017 51st Asilomar Conference on Signals, Systems, and
  Computers}, pages 1929--1934, 2017.

\bibitem{WunderEtAl:2018:Performance}
G.~Wunder, I.~Roth, R.~Fritschek, and J.~Eisert.
\newblock Performance of hierarchical sparse detectors for massive mtc.
\newblock {\em Preprint}, 2018.

\bibitem{wunder2018secure}
G.~Wunder, I.~Roth, R.~Fritschek, B.~Gro{\ss}, and J.~Eisert.
\newblock Secure massive iot using hierarchical fast blind deconvolution.
\newblock In {\em 2018 IEEE Wireless Communications and Networking Conference
  Workshops (WCNCW)}, pages 119--124. IEEE, 2018.

\bibitem{Wunder2019_TWC}
G.~{Wunder}, S.~{Stefanatos}, A.~{Flinth}, I.~{Roth}, and G.~{Caire}.
\newblock Low-overhead hierarchically-sparse channel estimation for multiuser
  wideband massive mimo.
\newblock {\em IEEE Transactions on Wireless Communications}, 18:2186--2199,
  April 2019.

\end{thebibliography}
    
\end{document}